\documentstyle{article}[12pt]

\begin{document}

\begin{center}
{\Large Optimal Renormalization-Group Improvement of $R(s)$ via the Method of Characteristics}
\end{center}

\bigskip

\begin{center}
V. {Elias $^{a,b,}$}\footnote{Electronic address: {\tt velias@uwo.ca}}, 
D.G.C. {McKeon $^{b}$}\footnote{Electronic address: {\tt dgmckeo2@uwo.ca}}
and T. G. {Steele $^{c}$}\footnote{Electronic address: {\tt steelet@sask.usask.ca}}
\end{center}

\baselineskip=12pt

\vskip .5cm

\begin{center}
{\small \it
$^{a}$ 
Perimeter Institute for Theoretical Physics, 35 King Street North,\\
Waterloo, Ontario  N2J 2W9 CANADA}\\
{\small \it
$^{b}$ 
Department of Applied Mathematics, The University of Western Ontario,\\
London, Ontario  N6A 5B7 CANADA}\\
{\small \it
$^{c}$ 
Department of Physics and Engineering Physics, University of  Saskatchewan,\\
Saskatoon, Saskatchewan  S7N 5C6  CANADA}
\end{center}

\begin{abstract}
We discuss the application of the method of characteristics to the 
renormalization-group equation for the perturbative QCD series within 
the electron-positron annihilation cross-section.  We demonstrate how
one such renormalization-group improvement of this series is equivalent
to a closed-form summation of the first four towers of 
renormalization-group accessible logarithms to all orders of 
perturbation theory.
\end{abstract}

\newpage

\setcounter{footnote}{0}

\baselineskip=24pt

The renormalization group (RG) has played a central role
in our understanding of quantum field theory [1-10] especially
since the discovery of asymptotic freedom [11-14].  The central
idea of the renormalization group is the insensitivity of physical
quantities to the mass scale $\mu^2$ introduced in the course of
regularizing and eliminating infinities within perturbative
calculations.  Explicit dependence of a perturbative series on $\mu^2$
is compensated by $\mu^2$ dependence in masses and coupling constants
characterising that series.  Indeed, the replacement of such quantities
by running quantities that are explicitly functions of $\mu^2$
constitutes what is generally denoted by ``RG-improvement'' of a
perturbative expression \cite{15}.  The numerical value of a calculation
to a given order of perturbation theory still depends upon the
numerical value of $\mu^2$, entailing the introduction of either prescriptions
(e.g., $m_b / 2 \leq \mu \leq 2m_b$ for semileptonic $b$-decays) or procedures
\cite{16,17} to obtain optimal values of $\mu^2$.

However, such substitutions do not in themselves take full advantage of
all information accessible from the renormalization-group equation
(RGE), which also determines portions of the perturbative series beyond
the order of perturbation theory to which calculations have been
explicitly performed.  Application of the RGE to one-loop expressions
has long been known to determine the leading logarithm contribution to
each subsequent order of perturbation theory.  The RGE can similarly be
used in conjunction with two-loop calculations to determine 
next-to-leading logarithm contributions to all subsequent orders in perturbation
theory --  indeed the application of the RGE to an $n^{th}$ loop
perturbative expression is sufficient to determine the contribution of
$n$ successively-subleading logarithms to all orders in the perturbative
expansion parameter.\footnote{The possibility that infrared effects
might alter this is addressed in ref. \cite{18}} 
Such RGE methods for obtaining and summing ``RG-accessible'' logarithms to 
all orders of perturbation theory have been applied to effective potentials
\cite{19} and actions \cite{20,
21}, QCD correlation functions \cite{22, 23}, QCD contributions to
decay rates \cite{22}, and even the high-energy behaviour of the $WW
\rightarrow ZZ$ cross-section \cite{22}, a process dominated by Higgs
boson exchanges. A related RG-summation of dimensionality poles
in the expansion of the bare coupling constant in terms of its
renormalized analog has been developed in ref. \cite{24} and (for
thermal field theory) in ref. \cite{25}.

The point we wish to emphasize is that the summation of higher order
logarithmic contributions is quite distinct (and a substantial
improvement over) what is usually
understood to be RG-improvement, the incorporation of running masses and
coupling constants into perturbative expressions taken to a given order.
Indeed, such inclusion of all RG-accessible logarithms within perturbative 
series is forcefully advocated in ref. \cite{26}.  Series which
incorporate summation of RG-accessible logarithmic
contributions to all orders of perturbation theory have been seen to
exhibit much less dependence on $\mu^2$ than series which utilize
running masses and coupling constants to a fixed calculational order
\cite{22, 23}.  This latter approach, however, devolves from the method
of characteristics \cite{27}, a standard approach to first-order partial
differential equations such as the RGE \cite{28}.  We demonstrate below how this
same method of characteristics can be extended to obtain summations of
leading and three successively-subleading towers of logarithms to all
orders of the perturbative  QCD series for the electron-positron
annihilation cross-section.

The total cross-section for $e^+ e^-$-annihilation, $R(s) \equiv \sigma
(e^+ e^- \rightarrow {\rm hadrons}) /$ $\sigma (e^+ e^- \rightarrow \mu^+
\mu^-)$ can be extracted from the imaginary part of the QCD 
vector-current correlation function \cite{29}, a perturbative expression 
that necessarily depends upon a renormalization mass scale $\mu$:
\begin{equation}
R(s) = 3 \sum_f Q_f^2 S \left[ x(\mu^2), \log (\mu^2 / s) \right].
\end{equation}
The expansion parameter $x(\mu^2) \equiv \alpha_s (\mu^2) / \pi$ is
proportional to the running QCD coupling constant, and renormalization
mass scale $\mu$ is a by-product of the regularization procedure for
indentifying and excising infinities from the underlying correlation
function, as discussed above.  Since $R(s)$ cannot depend on this
unphysical scale parameter, it follows that
\begin{equation}
\mu^2 \frac{d R(s)}{d\mu^2} = 0 = \left( \mu^2 \frac{\partial}{\partial
\mu^2} + \beta (x) \frac{\partial}{\partial x} \right) R(s)
\end{equation}
where
\begin{equation}
\beta(x(\mu^2)) \equiv \mu^2 \frac{dx(\mu^2)}{d\mu^2}
\end{equation}
with an appropriately chosen boundary condition for (3) [{\it e.g.}
$x(\Lambda^2) = \infty$ or $x (M_z^2) = 0.118 / \pi$].  Perturbative
series expansions of $S[x(\mu^2), \log \left( \frac{\mu^2}{s}\right)]$ and
$\beta(x)$ are seen to take the form

\renewcommand{\theequation}{4\alph{equation}}
\setcounter{equation}{0}

\begin{equation}
S[x,L] = 1 + \sum_{n=1}^\infty \sum_{m=0}^{n-1} T_{n,m} x^n L^m
\end{equation}
\begin{equation}
\beta(x) = -x^2 \sum_{k=0}^\infty \beta_k x^k
\end{equation}
where $x = x(\mu^2)$ and $L \equiv \log (\mu^2/s)$.  Generally $\beta(x)$ is determined by relating 
the bare coupling to the renormalized coupling \cite{7,9,24}, although it
can also be extracted directly from eq. (2) \cite{20,21,30}.  Indeed, explicit
Feynman diagrammatic calculations to four-loop order have determined
$\beta_0, \beta_1, \beta_2, \beta_3$ \cite{13, 14, 31} as well as $T_{1,0},
T_{2,1}, T_{2,0}, T_{3,2}, T_{3,1}$ and $T_{3,0}$ \cite{29}, and these
results (as tabulated in Table I of ref. [23]) are manifestly consistent with eq.
(2).

However, it is possible to utilize eq. (2) to extract higher-order
coefficients within $S[x,L]$ than those tabulated in Table 1.  It is
easily seen \cite{23} that $T_{1,0}$ and $\beta_0$ determine {\it all}
leading logarithm coefficients $T_{n, n-1}$ for $n > 1$;  similarly
additional knowledge of $T_{2,0}$ and $\beta_1$ is sufficient to
determine all next-to-leading logarithm coefficients $T_{n, n-2}$ for $n
> 2$;  knowledge of $T_{3,0}$ and $\beta_2$ permits determination of
$T_{n, n-3}$ for $n > 3$, and so forth. In ref. [23], the double
summation in eq. (4a) is reorganised into the form

\renewcommand{\theequation}{\arabic{equation}}
\setcounter{equation}{4}

\begin{equation}
S[x,L] = 1 + \sum_{n=1}^\infty x^n S_n (xL)
\end{equation}
where the functions
\begin{equation}
S_n (u) = \sum_{k=0}^\infty T_{n+k, k} u^k
\end{equation}
are completely determined by knowledge of the ``RG-accessible''
coefficients $T_{n+k, k}$.  One can show that eq. (2) gives rise to a
nested set of first order differential equations for the functions $S_n
(u)$:
\begin{equation}
\frac{dS_k}{du} - \frac{k\beta_0}{1-\beta_0 u} S_k = \frac{(1-
\delta_{k,1})}{1-\beta_0 u} \sum_{\ell = 1}^{k-1} \beta_\ell \left( u
\frac{d}{du} + k - \ell \right) S_{k-\ell} (u), \; \; S_n(0)=T_{n,0}.
\end{equation}
These equations are derived and sequentially solved in ref. [23]. When
one applies this ``RG-summation'' to the series (5) within $R(s)$, the
dependence of $R(s)$ on $\mu^2$ is substantially reduced \cite{22,23}.
This is not surprising, as the exact result for $R(s)$ is necessarily
independent of $\mu^2$ [the RGE is just a statement of this independence], 
and the inclusion of higher-order logarithm contributions to $R(s)$ via 
(5) is expected to approximate the exact result more closely than truncation 
of eq. (4) to a given order.

As discussed above, the method of characteristics [27,28] provides a
complementary procedure for obtaining information from the
renormalization group equation (2).  To illustrate this method, consider
the first-order partial differential equation
\begin{equation}
\left[ f(x,y) \frac{\partial}{\partial x} + g (x,y)
\frac{\partial}{\partial y} \right] A(x,y) = 0
\end{equation}
where $f$ and $g$ are given functions, and where $A(x,y)$ may be
indentified as some field-theoretical amplitude characterised by
quantities ({\it e.g}, coupling constants) $x$ and $y$.  If $A_0 (x,y)$ is a solution to eq.
(8), then so is $A_0 \left( \bar{x} (t), \bar{y} (t) \right)$, provided
that
\begin{equation}
\frac{d\bar{x}}{dt} = f \left( \bar{x} (t), \bar{y}(t) \right), \; \;
\frac{d\bar{y}}{dt} = g \left( \bar{x}(t), \bar{y}(t) \right)
\end{equation}
with initial conditions $\bar{x}(0) = x, \; \; \bar{y} (0) = y$.  One
then sees from eqs. (9) that
\begin{equation}
0 = \left[ f(\bar{x}, \bar{y}) \frac{\partial}{\partial \bar{x}} + g
(\bar{x}, \bar{y}) \frac{\partial}{\partial\bar{y}} \right] A_0
(\bar{x}, \bar{y}) = \frac{d}{dt} A_0 \left( \bar{x}(t), \bar{y}(t)
\right)
\end{equation}
The initial conditions ensure that $A_0 \left( \bar{x}(t), \bar{y}(t)
\right)$ is a solution of eq. (8) when $t = 0$.  Since eq. (10) implies
that $A_0 \left( \bar{x}(t), \bar{y}(t)
\right)$ is independent of $t$, $A_0 \left( \bar{x}(t), \bar{y}(t)
\right)$ is necessarily a solution to eq. (8) for all values of $t$.  
Eqs. (9) and (10) provide the justification for replacing the variables $x$ and
$y$ with their corresponding characteristic functions $\bar{x}, \bar{y}$ in the 
amplitude $A(x,y)$.

For the RGE (2), as applied to the field theoretical series $S[x, \log
(\mu^2 / s)]$, the role of $f$ and $g$ as dependent variables is assumed
by $\mu^2$ and $\beta$, in which case correspondence to eqs. (9) requires
running values for these functions
\begin{equation}
\frac{d\bar{\mu}^2 (t)}{dt} = \bar{\mu}^2 (t),
\end{equation}
\begin{equation}
\frac{d\bar{x} (t)}{dt} = \beta (\bar{x} (t)).
\end{equation}
The usual prescription for RG improvement is to identify $t$ with
$\log(\mu^2)$ [{\it i.e.}, $\bar{\mu}^2 = e^t = \mu^2$ via eq. (11)], in which
case eq. (12) becomes eq. (3).  Indeed, this construction provides the
justification for having the coupling constant $x$ run with $\mu^2$
within the perturbative series (4a) \cite{27}.

However, it is entirely valid to let $t$ be arbitrarily chosen in eqs.
(11) and (12), up to initial conditions $\bar{\mu}^2 (0) = \mu^2$,
$\bar{x}(0) = x(\mu^2)$ that establish contact with a known solution to
eq. (2).  Thus the ``running coupling'' $x(\mu^2)$ may be employed to
serve as an initial condition for the characteristic function $\bar{x}
(t)$.\footnote{The dimensional regularization equation relating the bare
$(g_B)$ and renormalized $(g)$ coupling constants is $g_B = \mu^\epsilon
\sum_{\ell=0}^\infty \sum_{k=\ell}^\infty a_{k,\ell} g^{2k+1}/\epsilon^\ell$
\cite{7,9}. Since $g_B$ is a bare parameter independent of $\mu$, the renormalized
coupling-constant $g$ is necessarily a $\mu$-dependent quantity, {\it i.e.}, 
a function of $\mu$.}

In order to keep track of the order of perturbation theory to which we
are working, we follow the approach of ref. [32] by introducing an
expansion parameter $\hbar$ such that

\renewcommand{\theequation}{13\alph{equation}}
\setcounter{equation}{0}

\begin{equation}
t \longrightarrow t/\hbar
\end{equation}
\begin{equation}
\bar{x} \longrightarrow \bar{x}\hbar
\end{equation}
so that the characteristic equations (9) become

\renewcommand{\theequation}{\arabic{equation}}
\setcounter{equation}{13}

\begin{equation}
\hbar \frac{d\bar{\mu}^2 (t)}{dt} = \bar{\mu}^2 (t)
\end{equation}
\begin{equation}
\hbar^2 \frac{d\bar{x} (t)}{dt} = - \bar{x}^2 \hbar^2 \sum_{n=0}^\infty \bar{x}^n \hbar^n \beta_n
\end{equation}
and the series expansion (4a) becomes
\begin{equation}
S[\bar{x}, \bar{L}] = 1 + \sum_{n=1}^\infty \sum_{m=0}^{n-1} T_{n,m} \hbar^n \bar{x}^n \bar{L}^m
\end{equation}
$[\bar{L} \equiv \log \left( \bar{\mu}^2 (t) / s \right)]$.  From eq. (14) and 
the $\bar{\mu}(0) = \mu$ initial condition, we see that 
\begin{equation}
\bar{\mu}^2 (t) = \mu^2 e^{t/\hbar}.
\end{equation}
We now express $\bar{x} (t)$ as a perturbative expansion
\begin{equation}
\bar{x}(t) = \sum_{n=0}^\infty \bar{x}_n (t) \hbar^n
\end{equation}
with $\bar{x}_0 (0) = x(\mu^2)$ and $\bar{x}_n (0) = 0$ for $n > 0$.  Upon subsituting eq. (18)
into eq. (15), we obtain a nested set of linear first-order differential equations for the 
variables $\bar{x}_n (t)$ when $n > 0$:
\begin{equation}
\frac{d\bar{x}_0}{dt} = - \beta_0 \bar{x}_0^2, \; \; \; \bar{x}_0 (0) =
x(\mu^2),
\end{equation}
\begin{equation}
\frac{d\bar{x}_1}{dt} + \left( 2\beta_0 \bar{x}_0 (t) \right) \bar{x}_1
= - \beta_1 \bar{x}_0^3 (t), \; \; \; \bar{x}_1 (0) = 0,
\end{equation}
\begin{equation}
\frac{d\bar{x}_2}{dt} + \left( 2\beta_0 \bar{x}_0 (t) \right) \bar{x}_2
= - \beta_2 \bar{x}_0^4 (t) - 3\beta_1 \bar{x}_0^2 (t) \bar{x}_1 (t) -
\beta_0 \bar{x}_1^2 (t), \; \; \; \bar{x}_2 (0) = 0,
\end{equation}
\begin{eqnarray}
\frac{d\bar{x}_3}{dt} + \left( 2\beta_0 \bar{x}_0 (t) \right) \bar{x}_3
= & - & \beta_3 \bar{x}_0^5 (t) - 4\beta_2 \bar{x}_0^3 (t) \bar{x}_1
(t)\nonumber\\
& - & 3\beta_1 \left(\bar{x}_0^2 (t) \bar{x}_2 (t) + \bar{x}_0 (t)
\bar{x}_1^2 (t) \right)\nonumber\\
& - & 2\beta_0 \bar{x}_1 (t) \bar{x}_2 (t), \; \; \; \bar{x}_3 (0) = 0.
\end{eqnarray}
The solution to eq. (19) is
\begin{equation}
\bar{x}_0 (t) = \frac{x(\mu^2)}{1 + \beta_0 x (\mu^2)t}.
\end{equation}
If we substitute eq. (23) into eq. (20), we find the solution to eq.
(20) to be
\begin{equation}
\bar{x}_1 (t) = -\frac{\beta_1}{\beta_0} \frac{x^2(\mu^2) \log (1 +
\beta_0 x (\mu^2) t)}{(1 + \beta_0 x(\mu^2) t)^2}.
\end{equation}
Similarly, subsitution of eq. (23) and (24) into eq. (21) leads to a
solution for $\bar{x}_2 (t)$
\begin{eqnarray}
\bar{x}_2 (t) & = & \frac{x^3(\mu^2)}{(1 + \beta_0 x (\mu^2) t)^3}
\left[ \left( \frac{\beta_1^2}{\beta_0^2} - \frac{\beta_2}{\beta_0}
\right) \beta_0 x(\mu^2) t - \frac{\beta_1^2}{\beta_0^2} \log (1+\beta_0
x (\mu^2) t)\right. \nonumber\\
& + & \left. \frac{\beta_1^2}{\beta_0^2} \log^2 (1 + \beta_0 x (\mu^2)
t) \right],
\end{eqnarray}
and substitution of eqs. (23), (24) and (25) into eq. (22) leads to a
solution of $\bar{x}_3 (t)$:
\begin{eqnarray}
\bar{x}_3 (t) & = & \frac{x^4(\mu^2)}{(1 + \beta_0 x(\mu^2) t)^4} \left[
\left( - \frac{\beta_1^3}{2\beta_0^3} + \frac{\beta_1 \beta_2}{\beta_0^2} -
\frac{\beta_3}{2\beta_0} \right) (1 + \beta_0 x(\mu^2) t)^2 \right.\nonumber\\
& + & \left( \frac{\beta_1^3}{\beta_0^3} - \frac{\beta_1
\beta_2}{\beta_0^2}\right) (1 + \beta_0 x(\mu^2) t) \left( 1 - 2\log (1
+ \beta_0 x(\mu^2) t) \right)\nonumber\\
& + & \left( -\frac{\beta_1^3}{2\beta_0^3} + \frac{\beta_3}{2\beta_0}
\right) + \left( \frac{2\beta_1^3}{\beta_0^3} - \frac{3\beta_1
\beta_2}{\beta_0^2} \right) \log \left( 1 + \beta_0 x(\mu^2)
t\right)\nonumber\\
& + & \left. \frac{5\beta_1^3}{2\beta_0^3} \log^2 \left( 1 + \beta_0 x(\mu^2)
t\right) - \frac{\beta_1^3}{\beta_0^3} \log^3 \left( 1 + \beta_0
x(\mu^2) t\right) \right].
\end{eqnarray}
(Eqs. (23), (24) and (25) also appear in refs. [21,32].)  Eqs. (23) -
(26) provide an expansion (18) of the solution to (15) that is distinct
from the usual perturbative expansion.

If we substitute the series (18) into the expansion (16) for
$S[\bar{x},\bar{L}]$ we find the following solution to the renormalization
group equation (2):
\begin{eqnarray}
S[\bar{x}, \bar{L}] & = & 1 + \hbar [T_{1,0} \bar{x}_0 (t)]\nonumber\\
& + & \hbar^2 \left[  T_{1,0} \bar{x}_1 (t) + (T_{2,0} + T_{2,1}
\bar{L}) \bar{x}_0^2 (t)]\right]\nonumber\\
& + & \hbar^3 \left[ T_{1,0} \bar{x}_2 (t) + \left( T_{2,0} + T_{2,1}
\bar{L}\right) \left( 2\bar{x}_0 (t)
\bar{x}_1(t)\right)\right.\nonumber\\
& & + \left. \left( T_{3,0} +T_{3,1} \bar{L} + T_{3,2} \bar{L}^2 \right)
\bar{x}_0^3 (t)\right]\nonumber\\
& + & \hbar^4 \left[ T_{1,0} \bar{x}_3 (t) + (T_{2,0} + T_{2,1}
\bar{L})(\bar{x}_1^2 (t) + 2\bar{x}_0 (t) \bar{x}_2
(t))\right.\nonumber\\
& & + \left( T_{3,0} + T_{3,1} \bar{L} + T_{3,2} \bar{L}^2 \right)
\left( 3\bar{x}_0^2(t) \bar{x}_1 (t) \right)\nonumber\\
& & + \left. \left( T_{4,0} + T_{4,1} \bar{L} + T_{4,2} \bar{L}^2 + T_{4,3}
\bar{L}^3 \right) \bar{x}_0^4 (t)\right]\nonumber\\
& + & ...
\end{eqnarray} 
Now, if $t = 0$, we find from eq. (17) that $\bar{\mu}^2 = \mu^2$, $\bar{L}
= \log(\mu^2/s) = L$.  Since $\bar{x}_0 (0) = x(\mu^2)$ and $\bar{x}_k
(0) = 0$ for $k \geq 1$, we see that eq. (27) recovers the original
expansion (4a) when $\hbar = 1$.

It $t$ is a non-zero constant, the solution (27) provides a means for
obtaining all coefficients $T_{n,m}$ with $m \neq 0$ in terms of
coefficients $T_{k,0}$.  To see this, let $t = \hbar \log k$, in which
case we see from eq. (17) that $\bar{\mu}^2 = k\mu^2$, $\bar{L} = \log
(k\mu^2/s) = L + \log k$.  If we substitute $t = \hbar \log k$ into eqs.
(23) - (26) and note that
\begin{eqnarray}
\log \left( 1 + \hbar \beta_0 x(\mu^2) t\right) & = & \hbar \beta_0
x(\mu^2)\log k\nonumber\\
& - & \frac{\hbar^2}{2} \beta_0^2 x^2 (\mu^2) \log^2 k\nonumber\\
& + & \frac{\hbar^3}{3} \beta_0^3 x^3 (\mu^2) \log^3 k + ... \; ,
\end{eqnarray}
we find upon further substitution into eq. (27) that
\begin{eqnarray}
S[\bar{x}, \bar{L}]|_{t = \hbar \log k} & = & 1 + \hbar x(\mu^2)
T_{1,0}\nonumber\\
& + & \hbar^2 x^2 (\mu^2) \left[ (T_{2,0} - \beta_0 T_{1,0} \log k) +
T_{2,1} \log (k\mu^2/s)\right]\nonumber\\
& + & \hbar^3 x^3(\mu^2) \left\{ \left[ T_{3,0} - (2 T_{2,0} \beta_0 +
T_{1,0} \beta_1) \log k + T_{1,0} \beta_0^2 \log^2 k \right]
\right.\nonumber\\
&& + \left. \left[ T_{3,1} - 2\beta_0 T_{2,1} \log k \right] \log
(k\mu^2/s) + T_{3,2} \log^2 (k\mu^2/s) \right\}\nonumber\\
& + & \hbar^4 x^4 (\mu^2) \left\{ \left[ T_{4,0} - \left( 3\beta_0 T_{3,0} +
2\beta_1 T_{2,0} + \beta_2 T_{1,0} \right) \log k \right.
\right. \nonumber\\
&& + \left. \left(3\beta_0^2 T_{2,0} + \frac{5}{2} \beta_0 \beta_1
T_{1,0} \right) \log^2 k - \beta_0^3 T_{1,0} \log^3 k \right]\nonumber\\
&& + \left[ T_{4,1} - (3 \beta_0 T_{3,1} + 2\beta_1 T_{2,1} ) \log k + 3
\beta_0^2 T_{2,1} \log^2 k \right] \log (k\mu^2/s)\nonumber\\
&& + \left. \left[ T_{4,2} - 3\beta_0 T_{3,2} \log k \right] \log^2
(k\mu^2/s) + T_{4,3} \log^3 (k\mu^2/s) \right\}\nonumber\\
&& + {\cal{O}} \left( \hbar^5 x^5 (\mu^2) \right).
\end{eqnarray}
Now if we rewrite the original series expansion (4a) with $L \equiv \log
(\mu^2/s) = \log (k\mu^2/s) - \log k$, we obtain
\begin{eqnarray}
S[x,L] & = & 1 + x(\mu^2) T_{1,0}\nonumber\\
& + & x^2 (\mu^2) \left[ (T_{2,0} - T_{2,1} \log k ) + T_{2,1} \log
(k\mu^2/s) \right]\nonumber\\
& + & x^3 (\mu^2) \left[ (T_{3,0} - T_{3,1} \log k + T_{3,2} \log^2 k)
\right. \nonumber\\
&& + (T_{3,1} - 2T_{3,2} \log k) \log (k\mu^2 / s) \nonumber\\
&& \left. + T_{3,2} \log^2 (k\mu^2/s)\right]\nonumber\\
& + & x^4 (\mu^2) \left[ (T_{4,0} - T_{4,1} \log k +  T_{4,2} \log^2 k -
T_{4,3} \log^3 k) \right.\nonumber\\
&& + (T_{4,1} - 2 T_{4,2} \log k + 3 T_{4,3} \log^2 k) \log
(k\mu^2/s)\nonumber\\
&& \left. + (T_{4,2} - 3 T_{4,3} \log k) \log^2 (k\mu^2/s) + T_{4,3} \log^3
(k\mu^2 / s) \right]\nonumber\\
& + & {\cal{O}} (x^5 (\mu^2)).
\end{eqnarray}

When $\hbar = 1$, eqs. (29) and (30) must be equal, since the solution (27)
to the renormalization-group equation (2) has been constructed...

\begin{itemize}
\item[1)]...to coincide with eq. (30) at $t = 0$, 
\item[2)]...to be independent of the choice for $t$ via the method of 
characteristics [eq. (10)], and
\item[3)]...to be given by eq. (29) when $t = \hbar \log k$.
\end{itemize}

\noindent Direct comparison of eqs. (29) and (30) when $\hbar = 1$ shows
that 
\begin{eqnarray}
T_{2,1} & = & \beta_0 T_{1,0},\nonumber\\
T_{3,1} & = & 2 T_{2,0} \beta_0 + \beta_1 T_{1,0},\nonumber\\
T_{3,2} & = & \beta_0^2 T_{1,0},\nonumber\\
T_{4,1} & = & 3\beta_0 T_{3,0} + 2\beta_1 T_{2,0} + \beta_2
T_{1,0},\nonumber\\
T_{4,2} & = & 3\beta_0^2 T_{2,0} + \frac{5}{2} \beta_0 \beta_1
T_{1,0},\nonumber\\
T_{4,3} & = & \beta_0^3  T_{1,0},
\end{eqnarray}
relations that can also be obtained  \cite{22} by direct substitution of
the series (4a) into the renormalization-group equation (2).  Thus, the
method of characteristics is seen to determine all logarithmic
coefficients to the order of perturbation theory considered.

However, a more powerful application of the solution (27) occurs by
setting $t = \hbar \log (s/\mu^2)$, ensuring via eq. (17) that
$\bar{\mu}^2 = s$, that $\bar{L} = 0$, and that factors of $1 + \hbar
\beta_0 x(\mu^2) t$ in eqs. (23) - (26) become $1 - \beta_0 x(\mu^2)
\log (\mu^2/s) \equiv w$ in the $\hbar \rightarrow 1$ limit.  In this
limit, eq. (27) generates the following series:
\begin{eqnarray}
S[\bar{x}, \bar{L}] & = & 1 + x(\mu^2) \frac{T_{1,0}}{w} +  x^2 (\mu^2) \left[ T_{2,0} - \frac{\beta_1}{\beta_0} T_{1,0} \log
w \right] w^{-2}\nonumber\\
& + & x^3 (\mu^2) \left[ T_{3,0} - 2T_{2,0} \frac{\beta_1}{\beta_0} \log
w  +  \left( \frac{\beta_1^2}{\beta_0^2} - \frac{\beta_2}{\beta_0}
\right) (w-1) - \frac{\beta_1^2}{\beta_0^2} \log w +
\frac{\beta_1^2}{\beta_0^2} \log^2 w \right]w^{-3}\nonumber\\
& + & x^4 (\mu^2) \left[ \left( - \frac{\beta_1^3}{2\beta_0^3} + \frac{\beta_1 \beta_2}{\beta_0^2} - \frac{\beta_3}{2\beta_0} \right) w^2
 + \left( \frac{\beta_1^3}{\beta_0^3} - \frac{\beta_1 \beta_2}{\beta_0^2} + 2T_{2,0} \left( \frac{\beta_1^2}{\beta_0^2} - \frac{\beta_2}{\beta_0} \right) \right) \right. w\nonumber\\
& + & \left( -\frac{2\beta_1^3}{\beta_0^3} + \frac{2\beta_1 \beta_2}{\beta_0^2} \right) w \log w + \left( -\frac{\beta_1^3}{2\beta_0^3} + \frac{\beta_3}{2\beta_0} 
-  2T_{2,0} \left( \frac{\beta_1^2}{\beta_0^2} -
\frac{\beta_2}{\beta_0}\right) + T_{4,0} \right) \nonumber\\
& + & \left( \frac{2\beta_1^3}{\beta_0^3} - \frac{3\beta_1 \beta_2}{\beta_0^2} 
-  2T_{2,0} \frac{\beta_1^2}{\beta_0^2} - 3T_{3,0} \frac{\beta_1}{\beta_0} \right) \log w \nonumber\\
& + & \left( \frac{5\beta_1^3}{2\beta_0^3} + 3T_{2,0} \frac{\beta_1^2}{\beta_0^2}\right) \log^2 w\nonumber\\
& - & \left. \frac{\beta_1^3}{\beta_0^3} \log^3 w \right] w^{-4} + \; ... 
\end{eqnarray}
This series explicitly reproduces the series (5) obtained via the
successive solutions to the differential equations (7).  The coefficient
functions $S_1 (xL)$, $S_2(xL)$, $S_3(xL)$ and $S_4(xL)$, as calculated
in ref. [23], are reproduced in eq. (32), demonstrating how the 
method-of-characteristics approach to the renormalization group equation can
recover the results obtained via summation of leading, next-to-leading,
and successively subleading logarithm factors to {\it all} orders of
perturbation theory.  Such results as eq. (32) represent the optimal
possible RG-improvement of the original perturbative series (4a),
insofar as they incorporate all RG-accessible coefficients of logarithms
occurring within that series.

As a final note, the ambiguity in the choice of $k$ such that $t = \hbar
\log k$ [thereby leading to the series of eq. (29)] is equivalent to the
ambiguity noted in ref. [25].  In \cite{25}, this ambiguity was viewed
as a consequence of shifting the initial condition of eq. (7) to the
equation (6) defining $S_n (u)$;  by replacing $\log (\mu^2 / s)$ with
$\log (k \mu^2 / s) - \log (k)$ in the series (4a), that series becomes
\begin{equation}
S[x,L] = 1 + \sum_{n=1}^\infty \sum_{m=0}^{n-1} T_{n, m}^\prime x^n (L^\prime)^m
\end{equation}
where $L^\prime \equiv \log (k \mu^2 / s)$ and where $T_{1,0}^\prime = T_{1,0} = 1$, $T_{2,1}^\prime = T_{2,1},
T_{2,0}^\prime = T_{2,0} - T_{2,1} \log (k)$, etc.  The initial condition in
eq. (7) is now replaced by $S_n (0) = T_{n,0}^\prime$.  Of course, when one
sums to all orders in perturbation theory, the dependence on $k$ within
eq. (29) will drop out.  

\section*{Acknowledgements}

We are grateful for research support from the Natural Sciences and
Engineering Research Council of Canada.  Discussions with A. Rebhan were
most helpful. V. E. would like to thank R. and D. MacKenzie for discussion.

\end{document}